\documentstyle[12pt]{article}

\columnwidth\textwidth

\begin{document}

\newcommand{\bda}{\begin{\displaymath}\begin{array}{rl}}
\newcommand{\eda}{\end{array}\end{displaymath}}
\newcommand{\be}{\begin{equation}}
\newcommand{\bdm}{\begin{displaymath}}
\newcommand{\edm}{\end{displaymath}}
\newcommand{\bea}{\begin{eqnarray}}
\newcommand{\eea}{\end{eqnarray}}
\newcommand{\bfm}{\boldmath}
\newcommand{\bfpi}{{\mbox{\boldmath{$\pi$}}}}
\newcommand{\bfphi}{{\mbox{\boldmath{$\phi$}}}}
\newcommand{\bftau}{{\mbox{\boldmath{$\tau$}}}}
\newcommand{\no}{\nonumber \\}
\newcommand{\fs}{\; \; .}
\newcommand{\co}{\; \; ,}
\newcommand{\per}{\;\;.}
\newcommand{\rar}{\rightarrow}
\newcommand{\nn}{\nonumber \\}
\newcommand{\mtiny}[1]{{\mbox{\tiny #1}}}
\newcommand{\al}{&\!\!\!\!}
\newcommand{\eff}{e\hspace{-0.1em}f\hspace{-0.18em}f}
\newcommand{\free}{\!f\hspace{-0.05em}r\hspace{-0.05em}e\hspace{-0.02em}e}
\newcommand{\ind}{\scriptscriptstyle}
\newcommand{\gra}{{\scriptscriptstyle\gamma}}
\newcommand{\QCD}{\mbox{\scriptsize Q\hspace{-0.1em}CD}}
\newcommand{\indR}{\mbox{\scriptsize R}}
\newcommand{\indL}{\mbox{\scriptsize L}}
\newcommand{\lvac}{\langle 0|\,}
\newcommand{\rvac}{\,|0\rangle}
\newcommand{\lav}{\langle\hspace{-0.2em}\langle}
\newcommand{\rav}{\rangle\hspace{-0.2em}\rangle}
\newcommand{\wave}{\raisebox{0.22em}{\fbox{\rule[0.15em]{0em}{0em}\,}}\,}
\newcommand{\mvec}{\vec{\rule{0em}{0.6em}m}}
\newcommand{\sm}{$\sigma$-model }
\newcounter{figuren}
\setcounter{figuren}{1}
\renewcommand{\thefigure}{\thefiguren}
\def\bibref[#1]{\cite{#1}}

\begin{titlepage}

\noindent
\hspace*{11cm} BUTP--94/8\\
\vspace*{1cm}
\begin{center}
{\LARGE Masses of the Light Quarks}
\footnote{Work supported in part by Schweizerischer Nationalfonds}

\vspace{3cm}

H. Leutwyler
\\
Institute for Theoretical Physics \\
University of Bern \\
Sidlerstrasse 5, CH-3012 Bern, Switzerland\\

{\scriptsize HLEUTWYLER@ITP.UNIBE.CH}
\\ \vspace{2cm}

Final version, December 1994 \\ \vspace*{1cm}

\nopagebreak[4]

\begin{abstract}
The magnitude of $m_u,\,m_d$ and $m_s$ is
discussed on the basis of Chiral Perturbation Theory. In particular, the
claim that $m_u=0$ leads to a coherent picture for the low energy structure
of QCD is examined in detail. It is pointed out
that this picture leads to violent flavour asymmetries in the matrix
elements
of the scalar and pseudoscalar operators, which are in conflict with the
hypothesis that the light quark masses may be treated as perturbations.
[Talk
delivered at the 2nd IFT Workshop, "Yukawa Couplings and the Origins of Mass",
Gainesville, Florida, Feb. 1994, to be published by International Press.]
\end{abstract}

\end{center}

\end{titlepage}

%%%%%%%%%%%%%%%%%%%%%%%%%%%%%%%%%%%%%%%%%%%%%%%%%%%%%%%%%%%%
\section{Effective low energy theory of QCD}
At low energies, the behaviour of scattering amplitudes or current matrix
elements can be described in terms of a {\it Taylor series expansion} in powers
of the momenta.
The electromagnetic form factor of the pion, e.g., may be
exanded in powers of the momentum transfer $t$.
In this case, the first two Taylor coefficients are related to the total charge
of the particle and to the mean square radius of the charge distribution,
respectively,
\be \label{taylor}
f_{\pi^+}(t) = 1 + \mbox{$\frac{1}{6}$} \langle r^2\rangle_{\pi^+}\, t +
O(t^2)\fs \end{equation}
Scattering lengths and effective ranges are analogous low energy
constants occurring in the Taylor series expansion of scattering amplitudes.

For the straightforward expansion in powers of the momenta to hold it is
essential that the theory does not contain massless particles. The exchange of
photons, e.g., gives rise to Coulomb scattering, described by an amplitude of
the form $e^2/(p'-p)^2$ which does not admit a Taylor series expansion. Now,
QCD does not contain massless particles, but it does contain very light ones:
pions. The occurrence of light particles gives rise to singularities in the low
energy domain which limit the range of validity of the Taylor series
representation. The form factor $f_{\pi^+}(t)$, e.g., contains a cut starting
at $t=4 M_\pi^2$, such that the formula (\ref{taylor}) provides an adequate
representation only for $t\ll 4 M_\pi^2$. To extend this representation to
larger momenta, one needs to account for the singularities generated by the
pions. This can be done, because the reason why $M_\pi$ is so small is
understood: the pions are the Goldstone bosons of a hidden, approximate
symmetry. The low energy singularities generated by the remaining members
of the pseudoscalar octet $(K^\pm, K^0, \bar{K}^0, \eta)$ can be dealt with in
the same manner, exploiting the fact that the Hamiltonian of QCD is
approximately invariant under $\mbox{SU(3)}_{\indR}
\times\mbox{SU(3)}_{\indL}$. If the three
light quark flavours $u,d,s,$ were massless, this symmetry would be an exact
one. In reality, chiral symmetry is broken by the quark mass term ocurring in
the QCD Hamiltonian
\bdm
H_{\QCD} =  H_0 + H_1  \co\hspace{2em}
H_1  =	\int\! d^3\!x \{m_u \bar{u}u + m_d \bar{d} d + m_s \bar{s}s \}\fs
\edm
For yet unknown reasons, the masses $m_u, m_d, m_s$ however happen to be small,
such that $H_1$ can be treated as a perturbation. First order perturbation
theory shows
that the expansion of the square of the pion mass in powers of $m_u, m_d, m_s$
starts with
\be\label{mpi}
M^2_{\pi^+} = (m_u + m_d) B \{ 1 + O (m_u, m_d, m_s) \}\co
\end{equation}
while, for the kaon, the leading term contains the mass of the strange quark,
\be\label{mk}
M^2_{K^+} = (m_u + m_s) B + \ldots \hspace{3em}
M^2_{K^0}  =  (m_d + m_s) B + \ldots
\end{equation}
This explains why the pseudoscalar octet contains the eight lightest hadrons
and why the mass pattern of this multiplet very strongly breaks eightfold way
symmetry: $M^2_\pi, M^2_K$ and $M^2_\eta$ are proportional to combinations of
quark masses, which are small but very different from one another, $m_s \gg m_d
> m_u$. For all other multiplets of SU(3), the main contribution to the mass is
given by the eigenvalue of $H_0$ and is of order $\Lambda_{\mbox{\tiny{QCD}}}$,
while $H_1$ merely generates a correction which splits the
multiplet, the state with the largest matrix element of $\bar{s}s$ ending up at
the top.

   The effective field theory combines the expansion in powers of momenta with
the expansion in powers of $m_u, m_d, m_s$. The resulting new improved Taylor
series, which explicitly accounts for the singularities generated by the
Goldstone bosons, is referred to as chiral perturbation theory
($\chi$PT).
It provides a solid mathematical basis for what used to be called the "PCAC
hypothesis" \bibref[Weinberg1979,GLAnnals,GLNP,found].

It does not appear to be possible to account for the singularities generated by
the next heavier bound states, the vector mesons, in an equally satisfactory
manner. The mass of the $\rho$-meson is of the order of the scale of QCD and
cannot consistently be treated as a small quantity. Although the vector meson
dominance hypothesis does lead to valid estimates (an example is given below),
a coherent framework which treats these estimates as leading terms of a
systematic approximation scheme is not in sight.

The effective low energy theory replaces the quark and gluon fields of QCD by a
set of pseudoscalar fields describing the degrees of freedom of the Goldstone
bosons $\pi, K, \eta$. It is convenient to collect these fields in a $3\times3$
matrix U$(x)\!\in\,$SU(3). Accordingly, the Lagrangian of QCD is replaced by an
effective Lagrangian, which only involves the field U$(x)$ and its derivatives.
The most remarkable point here is that this procedure does not mutilate the
theory: if the effective Lagrangian is chosen properly, the effective theory is
mathematically equivalent to QCD \bibref[Weinberg1979,found].

On the level of the effective Lagrangian, the combined expansion introduced
above amounts to an expansion in powers of derivatives and powers of the quark
mass matrix
\bdm
m = \left(
\begin{array}{ccc}
 m_u &	   &	 \cr
     & m_d &	 \cr
     &	   &  m_s
\end{array}
\right)
\edm
Lorentz invariance and chiral symmetry very strongly constrain the form of
the terms occurring in this expansion. Counting $m$ like two powers of
momenta, the expansion starts at $O(p^2)$ and only contains even terms
\bdm
{\cal L}_{\eff}= {\cal L}_{\eff}^2 +
{\cal L}_{\eff}^4 + {\cal L}_{\eff}^6 +
\ldots
\edm
The leading contribution is of the form
\be \label{Leff2}
{\cal L}_{\eff}^2 = \mbox{$\frac{1}{4}$}F^2 \mbox{tr} \{
\partial_\mu U^+ \partial^\mu U \} +
\mbox{$\frac{1}{2}$}F^2B \mbox{tr} \{ m (U+U^+) \}
\end{equation}
and involves two independent coupling constants: the pion decay constant
$F$ and the constant $B$ occurring in the mass formulae
(\ref{mpi}) and (\ref{mk}).
The expression (\ref{Leff2}) represents a compact summary of the soft pion
theorems
established in the 1960's: the leading terms in the chiral expansion of the
scattering amplitudes and current matrix elements are given by the tree graphs
of this Lagrangian.

At order $p^4$, the effective Lagrangian contains terms with four derivatives
such as
\bdm
{\cal L}_{\eff}^4 = L_1 [ \mbox{tr} \{ \partial_\mu U^+
\partial^\mu U \} ]^2 + \ldots
\edm
as well as terms with one or two powers of $m$. Altogether, ten coupling
constants occur \bibref[GLNP], denoted $L_1, \ldots, L_{10}$. Four of these are
needed
to specify the scattering matrix to first nonleading order. The terms of order
$m^2$ in the meson mass formulae (\ref{mpi}) and (\ref{mk}) involve
another three of
these constants. The remaining three couplings concern current matrix elements.

As an illustration, consider again the e.m. form factor $f_{\pi^+}(t)$. To
order $p^2$, the chiral representation reads \bibref[GLNP]
\be\label{f(t)}
f_{\pi^+} (t)  =
1 + \frac{t}{F^2} \{2L_9 + 2 \phi_\pi(t) + \phi_K(t)\}
+ O(t^2, t m)
\end{equation}
In this example, the leading term (tree graph of ${\cal
L}_{\eff}^2$) is trivial, because $f_{\pi^+}(0)$ represents the charge of
the particle. At order $p^2$, there are
two contributions: the term linear in $t$ arises from a tree graph of ${\cal
L}_{\eff}^4$ and involves the coupling constant $L_9$, while
the functions $\phi_\pi (t)$ and $\phi_K (t)$ originate in one loop graphs
generated by ${\cal L}_{\eff}^2$. The loop integrals contain
a logarithmic divergence which is absorbed in a renormalization of $L_9$ -- the
net result for $f_{\pi^+} (t)$ is independent of the regularization used. The
representation (\ref{f(t)}) shows how the straightforward Taylor series
(\ref{taylor}) is modified
by the singularites due to $\pi \pi$ and $K\bar{K}$ intermediate states. At the
order of the chiral expansion we are considering here, these singularities are
described by the one loop integrals $\phi_\pi(t), \phi_K(t)$ which contain
cuts starting at $t=4 M_\pi^2$ and $t = 4 M^2_K$, respectively. The result
(\ref{f(t)}) also shows that chiral symmetry does not determine the pion charge
radius: its
magnitude depends on the value of the coupling constant $L_9$ -- the effective
Lagrangian is consistent with chiral symmetry for any value of the coupling
constants.
The symmetry,
however, {\it relates} different observables. The slope of the $K_{l_3}$ form
factor $f_+(t)$, e.g., is also fixed by $L_9$. The experimental value of this
slope \bibref[PDG], $\lambda_+ = 0.030$, can therefore be used to first
determine
the magnitude of $L_9$ and then to calculate the pion charge radius. This gives
$\langle r^2\rangle_{\pi^+} = 0.42$ fm$^2$, to be compared with the
experimental result, 0.44 fm$^2$ \bibref[Amendolia].
%%%%%%%%%%%%%%%%%%%%%%%%%%%%%%%%%%%%%%%%%%%%%%%%%%%%%%%%%%%%%%%%%%%%%%%%%%
\section{Scale of the loop expansion}
One of the main problems encountered in the effective Lagrangian
approach is the occurrence of an entire fauna of effective coupling constants.
If these constants are treated as totally arbitrary parameters, the predictive
power of the method is equal to zero -- as a bare minimum, an estimate of their
order of magnitude is needed.

Let me first drop the masses of the light quarks and send the heavy ones to
infinity. In this limit, QCD is a theoretician's paradise: a theory without
adjustable dimensionless parameters. In particular, the effective coupling
constants $F, B, L_1, L_2, \ldots$ are
calculable --- the available, admittedly crude evaluations of $F$ and $B$ on
the lattice demonstrate that the calculation is even feasible. As discussed
above, the coupling constants $L_1, L_2, \ldots$ are renormalized by
the logarithmic divergences occurring in the one loop graphs. This property
sheds considerable light on the structure of the chiral expansion and provides
a rough estimate for the order of magnitude of the effective coupling
constants \bibref[Georgi]. The point is that the contributions generated by the
loop
graphs are smaller than the leading (tree graph) contribution only for momenta
in the range $\mid p \mid \raisebox{-0.3em}{$\stackrel{<}{\sim}$}\,
\Lambda_\chi$,
where \bibref[Soldate] \be\label{chiral scale}
\Lambda_\chi \equiv 4 \pi F/ \sqrt{N_f}
\end{equation}
is the scale occurring in the coefficient of the logarithmic divergence ($N_f$
is the number of light quark flavours). This indicates that the derivative
expansion is an expansion in powers of $(p/\Lambda_\chi)^2$ with coefficients
of order one. The stability argument also applies to the expansion in powers of
$m_u, m_d$ and $m_s$, indicating that the relevant expansion parameter is given
by $(M_\pi/ \Lambda_\chi)^2$ and $(M_K/\Lambda_\chi)^2$, respectively.
%%%%%%%%%%%%%%%%%%%%%%%%%%%%%%%%%%%%%%%%%%%%%%%%%%%%%%%%%%%%%%%%%%%%
\section{Low lying excited states}
A more quantitative picture can be obtained along the following lines. Consider
again the e.m. form factor of the pion and compare the chiral representation
(\ref{f(t)}) with the dispersion relation
\bdm
f_{\pi^+}(t) = \frac{1}{\pi} \int^{\infty}_{4M^2_\pi} \frac{dt'}{t'-t}
\mbox{Im} f_{\pi^+} (t')\fs
\edm
In this relation, the contributions $\phi_\pi, \phi_K$ from the one loop graphs
of $\chi$PT correspond to $\pi \pi$ and $K \bar{K}$ intermediate states. To
leading order in the chiral expansion, the corresponding imaginary parts are
slowly rising functions of $t$. The most prominent contribution on the r.h.s.,
however, stems from the region of the $\rho$-resonance which nearly saturates
the integral: the vector meson dominance formula, $f_{\pi^+} (t) =
(1-t/M_\rho^2)^{-1}$, which results if all other contributions are dropped,
provides a perfectly decent representation of the form factor for small values
of $t$. In particular, this formula predicts $\langle r^2\rangle_{\pi^+} =
0.39$ fm$^2$,
in satisfactory agreement with observation (0.44 fm$^2$). This implies that the
effective coupling constant $L_9$ is approximately given by \bibref[GLNP]
\be
L_9 = \frac{F^2}{2M^2_\rho}\fs
\end{equation}
In the channel under consideration, the pole due to $\rho$ exchange thus
represents the dominating low energy singularity -- the $\pi \pi$ and $K
\bar{K}$ cuts merely generate a small correction. More generally, the validity
of the vector meson dominance formula shows that, for the e.m. form factor, the
scale of the derivative expansion is set by $M_\rho = 770$ MeV.

Analogous estimates can be given for all effective coupling constants at order
$p^4$, saturating suitable dispersion relations with contributions from
resonances \bibref[Ecker,Leutwyler1990], e.g.
\be\label{L57}
L_5 = \frac{F^2}{4M^2_S}\co\hspace{3em}
L_7 = - \frac{F^2}{48M^2_{\eta'}}\co\end{equation}
where $M_S \simeq 980$ MeV and $M_{\eta'} = 958$ MeV are the masses of the
scalar octet and pseudoscalar singlet, respectively. In all those cases where
direct phenomenological information is available, these estimates do
remarkably well.
I conclude that the observed low energy structure is dominated
by the poles and cuts generated by the lightest particles -- hardly a surprise.
%%%%%%%%%%%%%%%%%%%%%%%%%%%%%%%%%%%%%%%%%%%%%%%%%%%%%%%%%%%
\section{Magnitude of the effective coupling constants}
The effective theory is constructed on the asymptotic states of QCD. In the
sector with zero baryon number, charm, beauty, $\ldots\,,$ the
Goldstone bosons form a complete set of such states, all other mesons
being unstable against decay into these (strictly speaking, the
$\eta$ occurs among the asymptotic states only for
$m_d=m_u$; it must be included among the degrees of freedom of the
effective theory, nevertheless, because the masses of the light quarks are
treated as a perturbation --- in massless QCD, the poles
generated by the exchange of this particle occur at $p=0$).
The Goldstone degrees of freedom are explicitly accounted for
in the effective theory --- they represent the dynamical variables. All other
levels manifest themselves only indirectly, through the values of the effective
coupling constants. In particular, low lying levels such as the $\rho$
generate relatively small energy denominators, giving rise to relatively large
contributions to some of these
coupling constants.

In some channels, the scale of the chiral expansion is set by $M_\rho$, in
others by the masses of the scalar or pseudoscalar resonances occurring around
1 GeV. This confirms the rough estimate (\ref{chiral scale}). The cuts
generated
by Goldstone pairs are significant in some cases and are negligible in others,
depending on the numerical value of the relevant Clebsch-Gordan coefficient. If
this coefficient turns out to be large, the coupling constant in question is
sensitive to the renormalization scale used in the loop graphs. The
corresponding pole dominance formula is then somewhat fuzzy, because the
prediction depends on how the resonance is split from the continuum underneath
it.

The quantitative estimates of the effective couplings given above
explain why
it is justified to treat $m_s$ as a perturbation. At order $p^4$, the symmetry
breaking part of the effective Lagrangian is determined by the
constants $L_4, \ldots, L_8$. These constants are immune to the low energy
singularities generated by spin 1 resonances, but are affected by the exchange
of scalar or pseudoscalar particles. Their magnitude is therefore determined
by the scale $M_S \simeq M_{\eta'} \simeq 1$ GeV. Accordingly,
the expansion in powers of $m_s$ is controlled by the parameter $(M_K/M_S)^2
\simeq \frac{1}{4}$. The asymmetry in the decay constants, e.g., is given
by \bibref[Leutwyler1990]
\be\label{fk/fpi}
\frac{F_K}{F_\pi} = 1+\frac{M^2_K - M^2_\pi}{M^2_S} + \chi\mbox{logs}
+O(m^2)\co \end{equation}
where the term $"\chi\mbox{logs}"$ stands for the chiral logarithms generated
by intermediate states with two Goldstone bosons.
This shows that the breaking of the chiral and eightfold way symmetries is
controlled by the mass ratio of the Goldstone bosons to the non-Goldstone
states of spin zero. In $\chi$PT, the observation that the Goldstones are the
lightest hadrons thus acquires quantitative significance.
%%%%%%%%%%%%%%%%%%%%%%%%%%%%%%%%%%%%%%%%%%%%%%%%%%%%%%%%%%%%%%%%%%%%
\section{Light quark masses}
A crude estimate for the order of magnitude of the light quark masses was
given shortly after the discovery of QCD
\bibref[GL74/75] :
\bdm
m_u \simeq 4\, \mbox{MeV}\co\hspace{2em} m_d \simeq 6
\,\mbox{MeV}\co\hspace{2em}  m_s \simeq
125-150 \,\mbox{MeV}\fs
\edm
Many papers dealing with the issue have appared since then. Weinberg
\bibref[Weinberg1977]
pointed out that $\chi$PT leads to an improved estimate for the ratios
$m_u:m_d:m_s$. Using the Dashen theorem \bibref[Dashen] to account for e.m.
self energies, the lowest order mass formulae given in eqs. (\ref{mpi}) and
(\ref{mk}) imply \bibref[Weinberg1977] \bea \label{ratiosalg}
\frac{m_u}{m_d} & =& \frac{M^2_{K^+}-  M^2_{K^0} + 2 M^2_{\pi^0} -
M^2_{\pi^+}} {M^2_{K^0} - M^2_{K^+} + M^2_{\pi^+}} \nonumber \\
\frac{m_s}{m_d} & = & \frac{M^2_{K^0} + M^2_{K^+} - M^2_{\pi^+}}
{M^2_{K^0} - M^2_{K^+} + M^2_{\pi^+}}\fs
\eea
Numerically, this gives
\be\label{ratiosnum}
\frac{m_u}{m_d} = 0.55 \co\hspace{3em}	\frac{m_s}{m_d} = 20\fs
\end{equation}
The higher order terms in the chiral expansion generate corrections
to the mass formulae (\ref{ratiosalg}), controlled by the parameter
$(M_K/\Lambda_\chi)^2$.
Roughly, the numerical result (\ref{ratiosnum}) should therefore hold to within
20 or 30\%.

The corrections of $O(p^4)$ to the above mass formulae
were worked out some time
ago \bibref[GLNP]. In particular, it was shown that these corrections drop out
when taking the double ratio
\be\label{defQ}
Q^2 \equiv \frac{M^2_K}{M_\pi^2}\cdot \frac{ M^2_K - M^2_\pi}{M^2_{K^0} -
M^2_{K^+}}\fs \end{equation}
The observed values of the meson masses thus provide a tight constraint on one
particular ratio of quark masses,
\be
Q^2 = \frac{m^2_s - \hat{m}^2}{m^2_d - m^2_u} \{ 1 + O (m^2) \}\co
\end{equation}
with $\hat{m} = \frac{1}{2} (m_u+m_d)$. The constraint may be visualized by
plotting the ratio
$m_s/m_d$ versus $m_u/m_d$ \bibref[KaplanManohar].
Dropping
the corrections of $O(m^2) = O(M_K^4/\Lambda_\chi^4)$, the
resulting curve takes the form of an ellipse,
\be\label{ellipse}
\left ( \frac{m_u}{m_d} \right)^2 + \,\frac{1}{Q^2} \left ( \frac{m_s}{m_d}
\right)^2 = 1\co
\end{equation}
with $Q$ as major semi-axis (the term $\hat{m}^2/m_s^2$ has been discarded,
as it is
numerically very small).
The meson masses occurring in the double ratio (\ref{defQ}) refer to pure QCD.
Using
the Dashen theorem to correct for the e.m. self energies, one
obtains $Q=24.1$.
For this value of the semi-axis, the ellipse passes through the point specified
by Weinberg's mass ratios, eq. (\ref{ratiosnum}).
%%%%%%%%%%%%%%%%%%%%%%%%%%%%%%%%%%%%%%%%%%%%%%%%%%%%%%%%%%%%%%%%%%%%%
\section{Corrections to the Dashen Theorem}
The Dashen theorem is subject to corrections from higher order terms in the
chiral expansion. As usual, there are two categories of contributions: loop
graphs of order $e^2m$ and terms of the same order from the derivative
expansion
of the effective e.m. Lagrangian.

The Clebsch-Gordan coefficients occurring in
the loop graphs are known to be large, indicating that two-particle
intermediate
states generate sizeable corrections; the
corresponding chiral logarithms
tend to increase the e.m. contribution to the kaon mass difference
\bibref[Langacker].
The numerical result depends on the scale used when evaluating the logarithms.
In fact, taken by themselves, chiral logs are unsafe at any scale.

The magnitude of the contributions from the terms of order $e^2m$
occurring in the effective Lagrangian is estimated in
two recent papers
by Donoghue et al. \bibref[DHW] and Bijnens \bibref[Bijnens]. Although the
framework used is rather different, they come up with the same conclusion: the
corrections are large, increasing the
value $(M_{K^+}-M_{K^0})_{e.m.} = 1.3$ MeV predicted by Dashen to $2.3$ MeV
(Donoghue et al.) and $2.6$ MeV (Bijnens), respectively.

In the present context, the main point is that even large corrections of this
size only lead to a small change in the value of $Q$, because the mass
difference between $K^+$ and $K^0$ is predominantly due to $m_d > m_u$ :
the value $Q = 24.1$ (Dashen) is lowered to $Q = 22.1$ (Donoghue et al.) and
$Q=21.6$ (Bijnens), respectively. Expressed in terms of the error bars attached
to the value of $Q$ in \bibref[Leutwyler1990], a doubling of the
electromagnetic self-energy modifies the value of $Q$ by
$1\frac{1}{2}\sigma$.

The decay $\eta \rightarrow 3\pi$ provides an independent measurement of $Q$:
writing the decay rate in the form $\Gamma_{\eta \rightarrow \pi^+\pi^-\pi^0} =
\Gamma_0/Q^4$, $\chi$PT predicts the value of $\Gamma_0$ in a parameter free
manner \bibref[GLNP]. Although the calculation accounts for all corrections of
$O(p^4)$, the numerical accuracy is rather modest because, due to strong final
state interaction effects, the calculated corrections are quite large.
Since the quantity $Q$ enters in the fourth power, the value $\Gamma_{\eta
\rightarrow \pi^+\pi^-\pi^0} = 283 \pm 28$ eV given by the particle
data
group \bibref[PDG] still yields a decent measurement: $Q = 20.6\pm 1.7$.
The fact that this result is significantly smaller than the value predicted
with the Dashen theorem represents an old puzzle.
The problem disappears if the e.m. contribution to the kaon mass difference
is significantly larger than indicated by the Dashen theorem
\bibref[DHW PRL,D,W].
In particular, the values of $Q$ obtained in \bibref[DHW,Bijnens] are
consistent with the one from $\eta$ decay.

The theoretical
uncertainties in the $\eta$ decay amplitude could be reduced. The calculation
referred to above only accounts for the corrections of order $p^4$ and
includes final state interaction effects and $\eta\eta'$ mixing
only to that order. A dispersive analysis along the lines indicated by Khuri
and Treiman \bibref[Khuri], which uses the $\chi$PT predictions only for the
subtraction
constants would likely lead to a more accurate estimate of the major
semi-axis \bibref[Anisovich].
%%%%%%%%%%%%%%%%%%%%%%%%%%%%%%%%%%%%%%%%%%%%%%%%%%%%%%%%%%%%%%%%%%%%%%%
\section{The ratio $m_u/m_d$}
Chiral perturbation theory thus fixes one of the two quark mass ratios in
terms of the other, to within small uncertainties. The ratios themselves,
i.e., the position on the ellipse, are a more subtle issue. Kaplan and
Manohar \bibref[KaplanManohar] pointed out that the
corrections to the lowest order result
(\ref{ratiosnum})
for $m_u/m_d$ cannot be determined on purely phenomenological grounds. They
argued that these corrections might be large and that the $u$-quark might
actually be massless. This possibility is of particular interest, because
the strong CP problem would then disappear. Several authors
\bibref[Seiberg] have given arguments in favour of $m_u = 0$.

Let me show the picture this reasoning leads to. The lowest order mass
formulae (\ref{mpi}), (\ref{mk}) imply that the ratio $m_u/m_d$ determines the
$K^0/K^+$ mass difference, the scale being set by $M_\pi$:
\be
M^2_{K^0} - M^2_{K^+} = \frac{m_d - m_u}{m_u + m_d}\cdot M^2_\pi + \ldots
\end{equation}
The formula holds up to corrections from higher order terms in the chiral
expansion and up to e.m. contributions. It is illustrated in figure
1, taken from \bibref[Dallas]. The
upper curve corresponds to the value $(M_{K^+}-M_{K^0})_{e.m.} =
1.3$ MeV, which follows
from the Dashen theorem. The correction of Donoghue et al. \bibref[DHW] shifts
the result
by 1 MeV (lower curve).
The horizontal line is the experimental value. Hence
the corrections from the higher order terms must generate the contributions
shown by the arrows. In particular, if $m_u$ is assumed to vanish, the lowest
order mass formula predicts a mass difference which exceeds the observed
value by a factor of four. The disaster can only be blamed on the "corrections"
from the higher order terms. It is evident that, under such circumstances,
it does not make sense to truncate the expansion at first nonleading order
and to fool around with the numerics of the effective coupling constants
occurring therein. The conclusion to draw from the assumption $m_u=0$ is that
$\chi$PT is unable to account for the masses of the Goldstone bosons. The fact
that it happens to work remarkably well in other cases must then be
accidental. I prefer to conclude that the assumption $m_u=0$ is not tenable.
\begin{figure}\vspace{2in}\caption{Sensitivity
of the first order prediction for the kaon
mass difference to $m_u/m_d$. The two curves
differ in the estimate used for the
e.m. self energies. E is the experimental value.
The dot corresponds to
Weinberg's leading order result.}
\end{figure}\addtocounter{figuren}{1}
\section{Phenomenological ambiguities}
\label{pheno}
The preceding presentation is an extract of a rapporteur talk given
in 1992 \bibref[Dallas].
I thought that this coffee was rather cold by now and dealt with it
only briefly in the manner described above, concentrating the remainder of the
talk
on issues which I expected to be more interesting to the audience, such as the
role of winding number and the mass of the $\eta^\prime$
\bibref[Smilga,Verbaarschot]. The vigorous discussion triggered by the
claim that
$m_u=0$ is firmly ruled out by now and the remarkable statements made by some
of the participants indicate, however, that not everyone fully shares this
view of the matter. I conclude that the arguments
given previously do not suffice and elaborate a little further.

To set up
notation, I first briefly review the observation of Kaplan and Manohar
\bibref[KaplanManohar], who pointed out that the matrix
\bdm
m' = \alpha_1 m + \alpha_2 (m^+)^{-1} \det m
\edm
transforms in the same manner as $m$.
For a real, diagonal mass matrix, the transformation amounts to
\be \label{KM1}
m_u' = \alpha_1 m_u + \alpha_2 m_d m_s \hspace{3mm} (\mbox{cycl.}\; u
\rightarrow d \rightarrow s \rightarrow u)\fs\end{equation}
Symmetry alone does therefore not distinguish $m'$ from $m$. If
${\cal L}_{\eff} (U, \partial U, \ldots , m)$ is an
effective Lagrangian consistent with chiral symmetry, so is
${\cal L}_{\eff} (U, \partial U, \ldots, m')$.

Since only the product $Bm$
enters the Lagrangian, $\alpha_1$ merely changes the value of the constant
$B$. The term proportional to $\alpha_2$ is a correction of order $m^2$
which, upon insertion in ${\cal L}^2_{\eff}$ generates a contribution to ${\cal
L}^4_{\eff}$. The contribution can again be removed by changing some of the
coupling constants:
\be\label{KM2}
B' = B/ \alpha_1 \co\;\; L'_6 = L_6 - \alpha\co\;\;
L'_7 =  L_7 - \alpha \co\;\; L'_8 = L_8 + 2 \alpha
\co\end{equation}
with $\alpha = \alpha_2 F^2/32 \alpha_1 B$. The effective Lagrangian is
therefore invariant under a simultaneous change of the quark mass matrix and of
the coupling constants.
Accordingly, the meson masses and scattering amplitudes
which one calculates with this Lagrangian remain the same.
The elliptic constraint considered above,
e.g., is invariant under the operation (up to terms of order
$(m_u - m_d)^2/m_s^2$,
which were neglected).
The
chiral representation
for the Green functions of the vector and axial currents are also invariant.
Since there is experimental information only about
masses, scattering amplitudes and matrix elements of the electromagnetic or
weak currents and since the chiral representation for these does not
distinguish
$m, B, L_i$ from $m', B', L_i'$, phenomenology does not allow one
to determine the magnitude of the constants $B, L_6, L_7, L_8$.

We are not dealing with a hidden symmetry of QCD here --- this
theory is not invariant under the change (\ref{KM1}) of the quark masses.
In particular, the matrix elements of the scalar and pseudoscalar operators
are modified.
Consider, e.g., the vacuum-to-pion matrix element of the pseudoscalar density.
The
Ward identity for the axial current implies that this matrix element is given
by \be\lvac\bar{d}\,i\gamma_5 u|\pi^+\rangle=\sqrt{2}\,F_{\pi^+}
M_{\pi^+}^2/(m_u+m_d)\end{equation}
The relation is exact, except for electroweak corrections. It involves the
physical quark masses and is not invariant under the above transformation.
For the ratio of the
$K$ and $\pi$ matrix elements, the effective Lagrangian yields the following
low energy representation:
\bdm\frac{\lvac\bar{s}\,i\gamma_5 u|K^+\rangle}{
\lvac\bar{d}\,i\gamma_5 u|\pi^+\rangle}=
1+\frac{4(M_{K^+}^2-M_{\pi^+}^2)}{F^2}(4
L_8-L_5)+\chi\mbox{logs}+O(m^2)\fs\edm
The relation is analogous to the formula for the ratio of the corresponding
matrix elements of the axial currents,
\bdm \frac{F_{K^+}}{F_{\pi^+}}=1 +\frac{4(M_{K^+}^2-M_{\pi^+}^2)}{F^2}\,L_5
+\chi\mbox{logs}+O(m^2)\fs\edm

While the ratio of decay constants is invariant under the KM transformation,
the one of the pseudoscalar densities is not.
The main difference between the two cases is that nature is kind
enough to provide us with a weak interaction probe, testing the matrix
elements of the vector and axial currents at low energies, while a probe which
would test
those of the scalar and pseudoscalar currents is not available --- the Higgs
particle is too heavy for this purpose.
The observed rates of the decays $K\rightarrow \mu\nu$ and $\pi\rightarrow
\mu\nu$ imply $F_{K^+}/F_{\pi^+}=1.22$ and thereby permit a phenomenological
determination of the
effective coupling constant $L_5$, while $L_8$ cannot be
determined on purely phenomenological grounds.

If the electroweak interactions were not available to probe the low energy
structure of QCD, the coupling
constant $L_5$ would also count amoung those quantities for which direct
phenomenological information is absent, for the following reason. The
field \be
U^\prime =U\{1+\alpha_3(U^\dagger m-m^\dagger U)-
\mbox{$\frac{1}{3}$}\alpha_3\mbox{tr}(U^\dagger m-m^\dagger U)
+O(m^2)\}\end{equation}
transforms in the same manner as $U$. For a constant
quark mass matrix, the change of variables $U\rightarrow U^\prime$
is equivalent to a change of the effective coupling constants:
\be\label{KM3} L_5^\prime =L_5-6\bar{\alpha}\co\;\;L_7^\prime=L_7 +
\bar{\alpha} \co\;\;L_8^\prime -3\bar{\alpha}
\co\;\;\bar{\alpha}=F^2\alpha_3/24 B\fs
\end{equation}
Hence the $\chi$PT results for the masses
of the Goldstone
bosons and for their scattering amplitudes only involve those combinations of
coupling constants which are invariant under this operation.
The only
difference to the transformation considered above is that (\ref{KM2}) leaves
the matrix elements of the vector and axial currents invariant, while
(\ref{KM3}) does not.

The ambiguity
pointed out by Kaplan and Manohar does not indicate that the effective
Lagrangian possesses any symmetries beyond those of the QCD Lagrangian.
It does not concern QCD as such, but originates in the
fact that the electromagnetic
and weak interactions happen to probe the low energy structure of the system
exclusively through
vector and axial currents.

%%%%%%%%%%%%%%%%%%%%%%%%%%%%%%%%%%%%%%%%%%%%%%%%%%%%%%%%%%%%%%%
\section{Additive renormalization of $m$
and all that}
One of the claims repeatedly made in the discussion is that the quark masses
occurring
in the effective Lagrangian must be distinguished from those entering the
Lagrangian of QCD.
Indeed, this problem does arise within
early formulations of the effective Lagrangian technique, which exclusively
dealt with the properties of the Goldstone bosons on the mass shell. There,
the structure of the effective Lagrangian was inferred from {\it global}
symmetry
arguments and the only place where the quark mass matrix entered was through
the transformation law of the term which explicitly breaks the symmetries
of the QCD Hamiltonian.

The
framework used in current work
on $\chi\mbox{PT}$, however, does not rely on
global symmetry arguments, but identifies the quark masses through their
contributions to the Ward identities, which express the symmetries of the
underlying theory in {\it local} form. The method was introduced
ten years ago \bibref[GLAnnals,found]. It is not limited to on-shell matrix
elements, but
also specifies the low energy expansion of the Green functions formed with the
vector, axial, scalar
and pseudoscalar currents.
The Ward identities
involve the physical quark masses, not some effective low
energy version thereof. {\it For the effective
theory to reproduce the low energy structure of the Green functions, the
mass matrix occurring in the effective Lagrangian must be identified
with the physical one}. Likewise, the magnitude of
the effective coupling
constants $L_6,\,L_7$ and $L_8$ is not an inherently ambiguous issue; the
effective theory reproduces the low energy expansion of QCD for precisely one
value of these constants.

There is a distinction between the currents associated with chiral
symmetry and the scalar or pseudoscalar densities in that the latter pick
up {\it multiplicative} renormalization, contragredient to the renormalization
of the quark mass matrix. It is crucial that the renormalization used is
mass independent, such that the Green functions of the scalar
and pseudoscalar operators are given by the response of the effective
action to a local change in the quark mass matrix.
In the ratio of matrix elements considered
above, the renormalization scale then drops out --- this ratio is a perfectly
well-defined pure number, which may be estimated, e.g., by means of QCD sum
rule techniques. In fact, the sum rule for the two-point functions of the
pseudoscalar currents
is one of the main sources of
information available today for the absolute magnitude of the light quark
masses. Once numerical evaluations on a lattice
reach sufficiently small quark masses, this method will allow a more precise
determination of low energy matrix elements involving the scalar and
pseudoscalar currents.

Some authors \bibref[Choi] have pointed out that {\it instantons} offer
a physical interpretation for the KM-ambiguity.
In the field of an
instanton, the
Dirac operator develops discrete zero modes. As pointed out by t'Hooft
\bibref[t'Hooft], these modes undergo an effective interaction which indeed
generates a self-energy
contribution for the up-quark proportional to $m_dm_s$. The problem
with this picture is
that the spectrum of the Dirac operator on which it is based
is fictitious. Banks and Casher \bibref[Banks Casher] have shown that the
spontaneous breakdown of chiral symmetry requires the small eigenvalues of the
Dirac operator to be distributed uniformly: the level
density $\rho(\lambda)$
approaches the value $\rho(0)= |\lvac\bar{q}q\rvac|/\pi$
when $\lambda\rightarrow 0$.
For the model described
in \bibref[Choi],
the spectrum instead contains an isolated zero mode,
such that the level density develops a peak there, $\rho(\lambda)\propto
\delta(\lambda)$. The peak may generate large
flavour symmetry breaking effects, but at the same time, it shows
that the model is in conflict with spontaneous chiral symmetry
breakdown. The calculations carried out in \bibref[Choi] are based on the
assumption that a framework which is unable to account for the leading term in
the effective Lagrangian (the one related to the constant
$B\leftrightarrow\lvac\bar{q}q\rvac$) can be trusted when analyzing the
higher order terms, i.e., the corrections.

There is some progress in understanding the level structure in
a picture which describes the vacuum as a dense collection of instantons
\bibref[Verbaarschot]. In that framework, the zero modes of the individual
instantons merge into a continuum, such that the fictitious peak mentioned
above disappears. For models with a decent
spectrum, the symmetry breaking effects should be of reasonable size.

%%%%%%%%%%%%%%%%%%%%%%%%%%%%%%%%%%%%%%%%%%%%%%%%%%%%%%%%%%%%%%%%%%%%%%%%%%%
\section{The $K^0-K^+$ mass difference}
As pointed out in \bibref[Dallas], the lowest order $\chi$PT formula
for the $K^0-K^+$ mass difference disagrees with observation by a factor of
four --- if $m_u$ is set equal to zero. Since I did not find this puzzle
discussed
in the literature \bibref[Seiberg], I do it myself and first try
to guess the reasoning put forward by the defence.

Any lawyer worth a fraction of his fee would immediately point out
that the problem
disappears if the quark masses occurring in the lowest order formula are
replaced by suitable KM-transforms
$m_u^\prime,\,m_d^\prime,m_s^\prime$. In particular, one may take $\alpha_1=1$
and
fix $\alpha_2$ in such a manner that the ratio of the primed masses agrees with
the Weinberg ratios. This amounts to the statement that the chiral
perturbation series should
be reordered, replacing the expansion in powers of the physical quark masses by
an expansion in the primed ones. With the above specification of these, the
leading term of the reordered series agrees with the experimental
value of
the $K^0-K^+$ mass difference, by construction.
So, the first conclusion to draw from $m_u=0$ is that the
expansion of the meson masses in powers of the physical quark masses does not
make sense --- $\chi$PT must be replaced by a reordered series.
Why does this not take care of the problem ?

The point is that the operation very strongly distorts the matrix
elements of the scalar operators, which are obtained from the
effective action by expanding in powers of the physical mass matrix, not
by some hand made variant thereof. Consider, e.g., the scalar form factors
\bdm \langle K^+|\bar{u}u-\bar{d}d|K^+\rangle=S_{K^+}(t)\co\hspace{2em}
\langle \pi^+|\bar{u}u-\bar{s}s|\pi^+\rangle=S_{\pi^+}(t)\fs\edm
Since the operation $s\leftrightarrow d$ takes a $K^+$ into a $\pi^+$, the
two
form factors coincide in the SU(3) limit. The difference
$S_{K^+}(t)-S_{\pi^+}(t)$ is a symmetry breaking effect of order $m_s-m_d$.
According to the Feynman-Hellman theorem \bibref[FeynmanHellman], the
values of these form factors at $t=0$ are related to the derivatives of the
meson masses with respect to the quark masses. The kaon
matrix element of the operator $\bar{u}u$, e.g., represents the first
derivative of $M_{K^+}^2$ with respect to $m_u$, such that
\bdm S_{K^+}(0)=\left(\frac{\partial}{\partial m_u}-
\frac{\partial}{\partial m_d}\right)M^2_{K^+}\co\hspace{2em}
S_{\pi^+}(0)=\left(\frac{\partial}{\partial m_u}-
\frac{\partial}{\partial m_s}\right)M^2_{\pi^+}\fs\edm
The expansion of the meson masses in powers of $m_u,m_d,m_s$
starts with \bibref[GLNP]
\bea M_{K^+}^2&=&(m_u+m_s)B\{1 +(m_u+m_s)K_3
+(m_u+m_d+m_s)K_4+\ldots\;\}\no
M_{\pi^+}^2&=&(m_u+m_d)B\{1 +(m_u+m_d)K_3
+(m_u+m_d+m_s)K_4+\ldots\;\}\co \nonumber\eea
where chiral logarithms and terms of $O(m^3)$ are dropped.
Evaluating the derivatives, one readily establishes the low energy theorem
\be r\equiv \frac{S_{K^+}(0)}
{S_{\pi^+}(0)}=\left(\frac{m_s-m_u}{m_d-m_u}
\cdot\frac{M_{K^0}^2-M_{K^+}^2}{M_{K^0}^2-M_{\pi^+}^2}
\right)^{\!\raisebox{-0.2em}{{\small 2}}}
\left \{\rule{0em}{1.2em}1 + O(m^2)\right\}\co\end{equation}
valid up to chiral logarithms (in the present case, these are proportional
to $M_\pi^2$ and therefore tiny, of order 1\%). The theorem is of the same
character as the
one which leads to the elliptic constraint: the corrections
are of second order in the quark masses.
The
relevant combination of quark masses which occurs here, however, fails to be
invariant under the
KM-transformation; the relation only holds if the masses entering the ratio
$(m_s-m_u)/(m_d-m_u)$ are identified with the physical ones.

Suppose now that $m_u$ vanishes. The formula then
predicts that the kaon
matrix element is about three times smaller than the pion matrix element
(the precise value of $r$
depends on the electromagnetic corrections: using the value
$(M_{K^0}-M_{K^+})_{\QCD}= 5.3\, \mbox{MeV}$, which follows from the Dashen
theorem, I obtain
$r=0.30$, while the values $(M_{K^0}-M_{K^+})_{\QCD}= 6.3\, \mbox{MeV}$
and $6.6\,\mbox{MeV}$ found by Donoghue et al. \bibref[DHW] and Bijnens
\bibref[Bijnens] lead to $r=0.36$ and $r=0.38$, respectively). So, $m_u=0$
leads
to the prediction that the evaluation of the above matrix elements with sum
rule or lattice techniques will reveal extraordinarily strong flavour symmetry
breaking effects.

Massless up-quarks usually come in a luxurious wrapping with plausible
well-known statements of general nature.
Indeed, there is
no logical contradiction
with the proposal that (i) the expansion
of the meson masses in powers of the physical quark masses fails and (ii)
a suitable reordering of the series leads to meaningful results.
Also, the prediction that the matrix elements
of the scalar and pseudoscalar operators exhibit dramatic
flavour symmetry
breaking effects is not in conflict with phenomenology. Although theoretical
tools such as sum rules
lead to the conclusion that the picture does not make sense,
these are more fragile than solid experimental facts. It
is conceivable that the present example represents an exception to Einstein's
statement: "Raffiniert ist der Herrgott, aber boshaft ist er nicht." Even
then, the literature on massless up-quarks leaves much to be desired, as it
does not address any of the consequences for the low energy structure of QCD,
which are rather bizarre.
%%%%%%%%%%%%%%%%%%%%%%%%%%%%%%%%%%%%%%%%%%%%%%%%%%%%%%%%%%%%%%%%%%%
\section{Does the quark condensate vanish ?}
Let me finally turn to an entirely different picture \bibref[Stern], which I do
not believe either, but find conceptually
more interesting. The picture may be motivated by an analogy with
spontaneous magnetization.
There, spontaneous symmetry breakdown
occurs in two quite different modes:
ferromagnets and antiferromagnets. For the former, the magnetization
develops a nonzero expectation value,
while for the latter, this does not
happen. In either case, the symmetry is spontaneosly broken (for a
discussion of the phenomenon within the effective Lagrangian framework,
see \bibref[Nonrel]). The example
illustrates that operators which the symmetry allows to pick up an expectation
value may, but need not do so.

The standard low energy analysis assumes
that the quark condensate is the leading order
parameter
of the spontaneously broken symmetry, such that the Goldstone boson masses are
proportional to the square root of the quark mass. The
proportionality coefficient follows from the relation of Gell-Mann,
Oakes and Renner \bibref[GMOR]
\bdm F_\pi^2  M_\pi^2=(m_u+m_d)|\lvac\bar{u}u\rvac|+\ldots\edm
The simplest version of the question
addressed in the papers
quoted above is whether the quark condensate indeed
tends to a nonzero limit when the quark masses are turned
off, like the magnetization of a ferromagnet, or whether it tends to zero, as
it is
the case for the magnetization of an antiferromagnet. If the second option were
realized in nature, the pion mass would not be determined by the quark
condensate, but by the terms of order $m^2$, which the
Gell-Mann-Oakes-Renner formula neglects. More generally, one may envisage a
situation, where the quark condensate is different from zero, but small, such
that, at the physical value of the quark masses, the terms of order $m$
and $m^2$ both yield a significant contribution. In the literature, this
framework is referred to as "Generalized $\chi$PT" \bibref[Stern]. The
"generalized" and
"ordinary" chiral perturbation theories only differ in the manner in which the
symmetry breaking effects are treated.

The main problem with G$\chi$PT is that much of the predictive power of
the standard framework is then lost. The most prominent example is the
Gell-Mann-Okubo formula for the octet of Goldstone bosons, which does not
follow within G$\chi$PT.
As mentioned above, this prediction of the standard framework is satisfied
remarkably well, thus supporting that picture, but it is not unfair to
say that this may be a coincidence, like the $\Delta I=\frac{1}{2}$-rule,
whose explanation within the Standard Model is also considerably more
complex than the rule itself.
The original motivation for the study of the generalized setting was the
discrepancy between theory and experiment with regard to the $\pi$N
$\sigma$-term, which indeed represented quite a fountain, continuously spitting
out new ideas, one more strange than the other. In the meantime,
this fountain has run dry,
because the discrepancy was shown to be due to an inadequate treatment of a
form factor, which enters the relation between the scattering amplitude and
the $\sigma$-term matrix element \bibref[GLS].
{}From my point of view, the generalized szenario is still of some interest,
mainly as a kinematical framework, which parametrizes the vicinity of the
$\chi$PT predictions and allows one to judge the
significance of the experimental information concerning these:
the standard framework represents a special case of the generalized one.
Although this is not very practical in general, one may even put
the cart before the horse and claim that the predictions
of $\chi$PT
are based on extra assumptions, which the generalized setting does not make
and which should be tested experimentally. Indeed, $\chi$PT leads to very
strong predictions concerning, e.g., the scattering lengths of $\pi\pi$
scattering \bibref[GLLett], which are sensitive to the explicit breaking
of chiral symmetry generated by the quark masses and
which it is important to test experimentally.
%%%%%%%%%%%%%%%%%%%%%%%%%%%%%%%%%%%%%%%%%%%%%%%%%%%%%%%%%%%%%%%%%%%%%%%
\section{Summary and Conclusion}

It is difficult for me to summarize the present knowledge concerning the
masses of the light quarks in an
unbiased manner. The following conclusions unavoidably reflect my own views and
mainly rely on work done together with Gasser.
These views are dyed with the
prejudice that a straightforward expansion of the matrix elements in powers of
the light quark masses
makes sense without reordering the series and that the Gell-Mann-Oakes-Renner
relation is not spoiled by higher order effects. \begin{figure}\vspace{2in}
\caption{The elliptic band shows the range permitted by the low
energy theorem (14). The cross-hatched area is the
intersection of this band with the sector selected by the phenomenology
of $\eta\eta^\prime$-mixing. The shaded
wedge shows the constraint imposed on the ratio $R$ by the mass splittings
in the baryon octet.}
\end{figure}

1. Concerning the absolute magnitude of the quark masses, the most reliable
estimates are based on evaluations of QCD sum rules \bibref[ITEP]. With the
experience gained from the various applications of these,
the evaluation leads to a remarkably stable result, the
visible noise generated by uncertainties of the input being quite
small \bibref[recentSR]. The
main uncertainty stems from the systematic error of the method,
which it is hard to narrow down. The results
do not indicate a significant change as compared to
the estimates given in \bibref[QM]. In the long run, lattice evaluations should
allow a more accurate determination, but further progress with light
dynamical fermions is required before the numbers obtained can be taken at
face value.

2. Chiral perturbation theory constrains the ratios $m_s/m_d$ and $m_u/m_d$ to
the ellipse specified in equation (\ref{ellipse}) and displayed in
figure 2 (taken from \bibref[Leutwyler1990]).
The magnitude of the semiaxis $Q$ is known to an accuracy of 10\%. The
value $Q= 24$ shown in the figure relies on the Dashen Theorem.
The estimates given in \bibref[DHW,Bijnens]
indicate that this theorem receives large corrections from
higher order terms, reducing the value of $Q$ by about 10\%.

3. The position on the ellipse
 is best discussed in terms of the ratio
\bdm R=\frac{m_s-\hat{m}}{m_d-m_u}\co\edm
which represents the relative magnitude of SU(3) breaking
compared to isospin breaking. If $m_u$ were to vanish, this ratio would be
related to the semiaxis by $R\simeq Q\simeq 24$ and the ratio of the strange
quark mass to $\hat{m}=\frac{1}{2}(m_u+m_d)$ would be given by
$m_s/\hat{m}\simeq 2Q\simeq 48$. The quark mass estimates obtained from
QCD sum rules and from numerical evaluations on a lattice
are in flat disagreement with this pattern,
but confirm the standard picture, where
$m_s/\hat{m}\simeq Q\simeq 24$.

4. The value of $R$ was investigated much before QCD had been
discovered. For a long time, it was taken for granted that isospin
breaking is an electromagnetic effect. The early literature on the subject is
reviewed in \bibref[QM].
In 1981, Gasser \bibref[Gasser1981] analyzed the
nonanalytic terms occurring in the chiral expansion of the octets of
meson and baryon
masses and demonstrated that these remove the discrepancies which had
affected previous determinations of the ratio $R$.
A thorough
discussion of
the available information concerning this ratio \bibref[QM] showed
that the data on
$\Sigma^+-\Sigma^-,\,\Xi^0-\Xi^-,\,K^0-K^+$ and $\rho\omega$-mixing
allow
four independent determinations of $R$ (in the case of the proton-neutron
mass difference,
the higher order corrections turn out to be large, so that this source of
information does not significantly affect the overall result). Within the noise
visible in the calculation, estimated at
10-15\%,
the results turned
out to be
mutually consistent. Adding the uncertainties
of the individual values quadratically, the analysis implies $R=43.5\pm
2.2$. Discarding the information from $K^0-K^+$ and from $\rho\omega$-mixing,
one instead obtains
$R=43.5\pm
3.2$, which corresponds to the range depicted as a shaded wedge in
figure 2.

5. The above error bars
rely on the prejudice that the matrix elements of the operators
$\bar{u}u,\,\bar{d}d,\,\bar{s}s$ do not exhibit strong SU(3) breaking
effects. These were estimated on the basis of the same
rough argument, which identifies
the quark mass difference $m_s-\hat{m}$ with the mass
splitting within the SU(3) multiplets \bibref[Lebed Luty].
When analyzing
the meson mass spectrum to second order in the chiral expansion \bibref[GLNP],
we realized, of course, that the mass spectrum of the Goldstone bosons does
not suffice to independently determine the ratios $Q$ and $R$, but the
algebraic nature of the ambiguity
was pointed out only later, by Kaplan and Manohar \bibref[KaplanManohar]. In
the opinion of some authors, this ambiguity appears
to represent the most interesting aspect of the matter, indicating that the
notion of quark mass is fuzzy at low energies, due to an {\it
additive
renormalization problem,} generated by {\it nonperturbative effects}, such as
{\it instantons}, etc.
The detailed discussion given
above is another attempt at fighting these dangerous ogres, which
the innocent observer might take for windmills (earlier
attempts are described in \bibref[Cervantes,Leutwyler1990,Dallas]).
Gradually, Rosinante is getting sick and tired.

6. One of the consequences we did infer from the second order analysis is
that the $\eta\eta^\prime$ mixing angle cannot be determined with
the Gell-Mann-Okubo formula ($\theta_{\eta\eta^\prime}=-
10^{\mbox{\footnotesize o}}$), because there
are other effects
of the same algebraic order of magnitude, originating in ${\cal L}_{\eff}^4$.
Using the information about the value of $R$ discussed above,
we obtained $\theta_{\eta\eta^\prime}=-20^{\mbox{\footnotesize o}}
\pm4^{\mbox{\footnotesize o}}$.
The phenomenological analysis of the decays $\eta\rightarrow\gamma\gamma$ and
$\eta^\prime\rightarrow\gamma\gamma$, performed independently a year
later \bibref[Donoghue Holstein Lin], neatly confirmed this prediction.

7. As noted in \bibref[GLNP], the result for the coupling constant $L_7$,
which follows from the above value of $R$, is consistent with the $\eta^\prime$
dominance formula (\ref{L57}). This
indicates that the sum rule used in the derivation of that formula is nearly
saturated by the $\eta^\prime$ intermediate state also in the real
world, not only in the large $N_c$ limit, where the formula is exact.

8. Turning the argument
around, the phenomenology on the mixing angle available today
\bibref[mixing angle]
may be used to
determine the quark mass
ratios. This is indicated by the dashed straight lines in figure 2.
Since the data agree with the prediction,
it is clear that the information about the mass ratios obtained in this manner
is consistent with the
one extracted from isospin breaking in the baryon octet. Note, however, that
the phenomenological analysis of the $\eta$ and
$\eta^\prime$ decays
relies on large $N_c$ arguments. Although the
resulting picture yields a coherent
understanding of quite a few processes \bibref[Bijnens Bramon Cornet,Dafne],
the validity of these arguments needs to be
examined more carefully to arrive at firm conclusions \bibref[Shore
Veneziano].

9. The branching ratio $\Gamma(\psi^\prime\rightarrow\psi\pi^0)/
\Gamma(\psi^\prime\rightarrow\psi\eta)$ provides further information about
$R$. Using the lowest order formula, the data available at the time gave
$R=
28\,$\raisebox{-0.2em}{$\stackrel{\mbox{\scriptsize +7}}{\mbox{\scriptsize
$- 4$}}$} \bibref[QM].
The formula is valid up to SU(3) breaking effects;
according to the general rule of thumb, these are expected to be
of order 20 or 30\%. Since we saw no way of evaluating these, the information
derived from this branching ratio did not significantly affect our analysis and
was discarded.
Meanwhile, the data are slightly more
precise \bibref[PDG]; the lowest order formula now yields
$R=31\pm4$. Furthermore, Donoghue and Wyler \bibref[DW] have investigated the
second order corrections, using the multipole expansion, which relates the
relevant matrix elements to $\lvac
G_{\mu\nu}\tilde{G}^{\mu\nu}|\pi^0\rangle$, $\lvac G_{\mu\nu}\tilde{G}^{\mu\nu}
|\eta\rangle$. As it is the case with the sum rule used to estimate $L_7$,
these matrix elements involve pseudoscalar operators and do thus not suffer
from the KM-ambiguity.
The calculation yields remarkably small SU(3)-breaking
effects, such that the value obtained for $R$ remains close to $31\pm4$.

The problems faced by this
attempt at estimating $R$ are listed in
\bibref[Dallas]. In particular, as discussed in detail in \bibref[Luty], it is
not clear that the relevant matrix elements may
be replaced by those of the operator $G_{\mu\nu}\tilde{G}^{\mu\nu}.$
The calculation is of interest, because it is independent of other
determinations, but at the present level of theoretical understanding,
it is subject to considerable uncertainties. In view of these,
it appears to me that the numerical result is perfectly consistent with
the value $R=43.4$, which follows from the Weinberg ratios.

As evidenced by the example of $\eta\eta^\prime$ mixing and quite a
few others,
the estimates of the effective coupling constants which follow from the
hypothesis that the low energy structure is dominated by the singularities due
to the lowest lying levels lead to a rather predictive
framework, which until now has passed all experimental tests.
It would be of interest
to estimate the symmetry breaking effects in the branching ratio
$\Gamma(\psi^\prime\rightarrow\psi\pi^0)/
\Gamma(\psi^\prime\rightarrow\psi\eta)$
on the basis of this hypothesis and
to see whether they indeed
increase the value of $R$ obtained with the lowest order formula. If not,
there would be a problem with the simple picture I am advocating here.

10. There are examples,
where the higher order corrections turn out to be large, such as
$\eta\rightarrow 3\pi$ or scalar form factors. In all cases I know of,
one can put the finger on the culprit responsible for the
enhancement. Invariably, the
problem arises, because the perturbations generated by the quark mass term
are enhanced by small energy denominators. In particular, the enhancement due
to strong final state interactions in the S-waves often generates sizeable
corrections, which are perfectly well understood ---
$\chi$PT itself predicts how large the S-wave phase shifts
are near threshold and that they rapidly grow with energy.
Once the origin of the phenomenon is understood, one may
take it into account and arrive at a reliable low energy representation,
even if the straightforward chiral expansion thereof contains relatively
large contributions from higher order terms.

11. In the case of the mass formulae, the low energy singularities do not
generate a significant
enhancement of the higher order contributions. Accordingly, the
estimates of the effective
coupling constants mentioned above
imply that the Weinberg ratios only receive small corrections.
There is some
evidence for a 10\% decrease in $Q$,
related to the e.m.
self energies:
The Dashen theorem is suspected to receive large corrections,
because some of the low energy singularities do generate
large symmetry breaking effects in that case.

12. Although the above hypothesis
is the most natural setting for an effective theory I can think of,
it evidently involves assumptions which go beyond pure
symmetry and phenomenology.
One may dismiss this hypothesis and the estimates for $R$ obtained
with
it, requiring only that the mass term of the light quarks, which breaks the
chiral symmetry of the QCD Hamiltonian, may be treated as a
perturbation.
With $m_u=0$, the lowest order formula for $M_{K^0}-M_{K^+}$ is
off by a factor of four, in flat contradiction with this requirement.
The disaster can be avoided only if
the factor $(m_d-m_u)/(m_d+m_u)$ accounts for at least half of the
discrepancy, $(m_d-m_u)/(m_d+m_u)<
\frac{1}{2}$. This alone yields $m_u/m_d>\frac{1}{3}$.

13. Accordingly, if $m_u$ is assumed to be equal to zero, $\chi$PT
must be thrown overboard. The expansion in powers of
the quark masses does then not make sense and
the matrix elements
of the scalar and pseudoscalar
operators must exhibit extraordinarily strong SU(3) breaking effects.
These cannot be explained with the
low energy singularities listed in the particle data
tables. As far as I know,
the instanton model
of Choi et al. \bibref[Choi] is the only theoretical scenario proposed to
substantiate the claim that QCD may give rise to
strong flavour symmetry breakings of this type.
There, the occurrence of discrete zero modes does indeed produce such effects.
The same feature, however, is the reason why
the model is inconsistent with spontaneously broken
chiral symmetry.
Dilute instanton coffee is undrinkable. Bold discussion remarks may warm it
up a little, but that merely intensifies the unacceptable flavour
characteristics of the brewing.\\

\noindent I am indebted to Hans Bijnens, J\"{u}rg Gasser, Marina Nielsen and
Daniel Wyler for valuable comments.
%%%%%%%%%%%%%%%%%%%%%%%%%%%%%%%%%%%%%%%%%%%%%%%%%%%%%%%%%%%%%%%%%%%%


\begin{thebibliography}{99}
\bibitem{Weinberg1979}{S. Weinberg, {\it Physica} A96 (1979) 327.}

\bibitem{GLAnnals}{J. Gasser and H. Leutwyler, {\it Ann. Phys. (N.Y.)\/}
158 (1984) 142.        }

\bibitem{GLNP}{J. Gasser and H. Leutwyler,
{\it Nucl. Phys.\/} B250 (1985) 465, 517, 539.}

\bibitem{found}{The foundations of the method
are discussed in detail in \\
H. Leutwyler, {\it Ann. Phys. (N.Y.)} 235 (1994) 165.}

\bibitem{PDG}{Review of Particle Properties, {\it Phys. Rev.} D45 (1992).}

\bibitem{Amendolia}{S.R. Amendolia et al., {\it Nucl. Phys.} B277 (1986) 168.}

\bibitem{Georgi}{H. Georgi, {\it Weak Interactions and Modern Particle Theory}
\\(Benjamin/Cummings, Menlo Park, 1984);
\\
H. Georgi and A. Manohar, {\it Nucl. Phys.} B234 (1984) 189.}

\bibitem{Soldate}{M. Soldate and R. Sundrum, {\it Nucl. Phys.} B340 (1990) 1;
\\
R.S. Chivukula, M.J. Dugan and M. Golden, {\it Phys. Rev.} D47 (1993) 2930.}

\bibitem{Ecker}{G. Ecker et al., {\it Nucl. Phys.} B321 (1989) 311; {\it Phys.
Lett.} B223 (1989) 425.}

\bibitem{Leutwyler1990}{H. Leutwyler, {\it Nucl. Phys.\/} B337 (1990)
108.}

\bibitem{GL74/75}{H. Leutwyler, {\it Phys. Lett.} B48 (1974) 431; {\it Nucl.
Phys.} B76 (1974) 413;
\\
J. Gasser and H. Leutwyler, {\it Nucl. Phys.} B94 (1975) 269.}

\bibitem{Weinberg1977}{S. Weinberg, in {\it A Festschrift for I.I. Rabi}, ed.
L. Motz (New York Acad. Sci., 1977) p. 185.	}

\bibitem{Dashen}{R. Dashen, {\it Phys. Rev.} 183 (1969) 1245.}

\bibitem{KaplanManohar}{D. B. Kaplan and A. V. Manohar, {\it Phys. Rev.
Lett.\/} 56 (1986) 2004.	}

\bibitem{Langacker}{P. Langacker and H. Pagels, {\it Phys. Rev.} D8 (1973)
4620;
\\
K. Maltman and D. Kotchan, {\it Mod. Phys. Lett.} A5 (1990) 2457;
\\
G. Stephenson, K. Maltman and T. Goldman, {\it Phys. Rev.}
D43 (1991) 860.}

\bibitem{DHW}{J. Donoghue, B. Holstein and D. Wyler, {\it Phys. Rev.} D47
(1993) 2089.}

\bibitem{Bijnens}{J. Bijnens, {\it Phys. Lett.} B306 (1993) 343.}

\bibitem{DHW PRL}{J. Donoghue, B. Holsten and D. Wyler, {\it Phys. Rev. Lett.}
69 (1992) 3444.}

\bibitem{D}{J. Donoghue, Lectures given at the Theoretical Advanced Study
Institute (TASI), Boulder, Colorado (1993).}

\bibitem{W}{D. Wyler, Proc. XVI Kazimierz Meeting on Elementary Particle
Physics, eds. Z. Ajduk et al., World Scientific (1994).}

\bibitem{Khuri}{N. Khuri and S. Treiman, {\it Phys. Rev.} 119 (1960) 1115;
\\
C. Roiesnel and T.N. Truong, {\it Nucl. Phys.} B187 (1981) 293.}

\bibitem{Anisovich}{A. V. Anisovich, "Dispersion relation technique for
three-pion system and the P-wave interaction in $\eta\rightarrow 3\pi$
decay", preprint Petersburg Nuclear Physics Institute, Gatchina
TH-62-1993/1931;\\ J. Kambor, C. Wiesendanger and D. Wyler, in preparation;\\
A. V. Anisovich and H. Leutwyler, in preparation.}

\bibitem{Seiberg}{For a review, see T. Banks, Y. Nir and N. Seiberg, in these
Proceedings.}

\bibitem{Dallas}{H. Leutwyler, Proc. XXVI Int. Conf. on High Energy Physics,
Dallas, Aug. 1992, ed. J. R. Sanford, AIP Conf. Proc. No. 272, (1993).}

\bibitem{Smilga}{H. Leutwyler and A. Smilga,
{\it Phys. Rev.\/} D 46 (1992) 5607;\\
A. Smilga and J. Stern, {\it Phys. Lett.} B318 (1993) 531.}

\bibitem{Verbaarschot}{E. Shuryak and J. Verbaarschot, {\it Nucl. Phys.} B 410
(1993) 37;\\ E. Shuryak and J. Verbaarschot, {\it Nucl. Phys.}
A 560 (1993) 306;\\
J. Verbaarschot and I. Zahed,
{\it Phys. Rev. Lett.} 70 (1993) 3852;\\
J. Verbaarschot, {\it Nucl. Phys.} B 426 (1994) 559; B 427 (1994) 534;\\ {\it
Phys. Rev. Lett.} 72 (1994) 2531; {\it Phys. Lett.} B 329 (1994) 351;\\
A. Smilga and J. Verbaarschot, "Spectral sum sules and finite volume partition
function in gauge theories with real and pseudoreal fermions", preprint Univ.
Minnesota, TPI-MINN-94/10-T (1994).
}

\bibitem{Choi}{K. Choi, C.W. Kim and W.K. Sze, {\it Phys. Rev. Lett.\/} 61
(1988) 794; \\
K. Choi and C.W. Kim, {\it Phys. Rev.\/} D40 (1989) 890;
\\
K. Choi, {\it Nucl. Phys.\/} B383 (1992) 58; {\it Phys. Lett.\/} B292 (1992)
159.}

\bibitem{t'Hooft}{G. t'Hooft, {\it Phys. Rev.} D14 (1976) 3432.}

\bibitem{Banks Casher}{T. Banks and A. Casher, {\it Nucl. Phys.} B169 (1980)
103.}

\bibitem{FeynmanHellman}{H. Hellman, "Einf\"{u}hrung in die Quantenchemie",
Deuticke, Leipzig (1937);\\
R. P. Feynman, {\it Phys. Rev.} 66 (1939) 340.}

\bibitem{Stern}{For a recent exposition of G$\chi$PT, see\\
J. Stern, H. Sazdjan and N. H. Fuchs, {\it Phys. Rev.} D47 (1993) 3814;\\M.
Knecht et al., {\it Phys. Lett.} B313 (1993) 229.\\
The early literature is reviewed in
\\ M. D. Scadron, Rep. Prog. Phys. 44 (1981) 213.}

\bibitem{Nonrel}{S. Randjbar-Daemi, A. Salam and
J. Strathdee, {\it Phys. Rev.} B48 (1993) 3190;
\\
H. Leutwyler, {\it Phys. Rev.} D49 (1994) 3033.}

\bibitem{GMOR}{M. Gell-Mann, R. J. Oakes and B. Renner, {\it Phys. Rev.} 175
(1968) 2195.}

\bibitem{GLS}{J. Gasser, H. Leutwyler and M. Sainio, {\it Phys. Lett.} B253
(1991) 260.}

\bibitem{GLLett}{J. Gasser and H. Leutwyler, {\it Phys. Lett.} B125 (1983) 321,
325.}

\bibitem{ITEP}{A. I. Vainshtein et al., {\it  Sov. J. Nucl. Phys.} 27
(1978) 274; \\
B. L. Ioffe, {\it Nucl. Phys.} B188 (1981) 317; B191 (1981) 591(E).}

\bibitem{recentSR}{The literature may be traced with the following
papers:\\
C. Adami, E. G. Drukarev and B. L. Ioffe {\it Phys. Rev.} D48 (1993) 2304;\\
V. L. Eletsky and B. L. Ioffe, {\it Phys. Rev.} D48 (1993)
1441;\\
C. A. Dominguez, C. Van Gend and N. Paver, {\it Phys. Lett.} B253
(1991) 241;\\
S. Narison, {\it Phys. Lett.} B216 (1989) 191;\\
C. A. Dominguez and E. de Rafael, {\it Ann. Phys.} 174 (1986) 372;\\
X. Jin, M. Nielsen and J. Pasupathy, "Calculation of $\langle
p\,\rule{.02em}{.7em}\,\bar{u}u\!-\!\bar{d}d\,\rule{.02em}{.7em}\,p\rangle$
from
QCD sum rule and the neutron-proton mass difference", hep-ph/9405202.}

\bibitem{QM}{J. Gasser and H. Leutwyler, {\it Phys. Reports} 87 (1982) 77.}

\bibitem{Gasser1981}{J. Gasser, {\it Ann. Phys. (N.Y.)} 136 (1981) 62.}

\bibitem{Lebed Luty}{The chiral logarithms occurring
in the expansion are worked out in\\
R. F. Lebed and M. A. Luty, {\it Phys. Lett.} B 329 (1994) 479.}

\bibitem{Cervantes}{M. de Cervantes, "Don Quixote", part I, chapter 8, Madrid
(1605).}

\bibitem{Donoghue Holstein Lin}{J. F. Donoghue, B. R. Holstein
and Y. C. R. Lin,
{\it Phys. Rev. Lett.} 55 (1985) 2766.}

\bibitem{mixing angle}{F. Gilman and R. Kauffmann, {\it Phys. Rev.}
D36 (1987) 2761;\\
Riazuddin and Fayyazuddin, {\it Phys. Rev.} D37 (1988) 149;\\
ASP Collaboration, N. A. Roe et al., {\it Phys. Rev.} D41 (1990) 17.}

\bibitem{Bijnens Bramon Cornet}{J. Bijnens, A. Bramon and F. Cornet, {\it Z.
Phys.} C46 (1990) 599.}

\bibitem{Dafne}{The DAFNE Physics Handbook, eds. L. Maiani, G. Pancheri and N.
Paver, INFN--Frascati (1992).}
\bibitem{Shore Veneziano}{G. M. Shore and G. Veneziano, {\it Nucl. Phys.} B381
(1992) 3.}

\bibitem{DW}{J. Donoghue and D. Wyler, {\it Phys. Rev.} D45 (1992) 892.}

\bibitem{Luty}{K. Gottfried, {\it Phys. Rev. Lett.} 40 (1978) 598;\\
Y. P. Tung and T. N. Yan, {\it Phys. Rev.} D41 (1990) 155;\\
M. Luty and R. Sundrum, {\it Phys. Lett.} B312 (1993) 205.}

\end{thebibliography}
\end{document}